\documentclass[10pt]{iopart}
\usepackage{graphicx}
\usepackage{amsfonts}
\usepackage{amssymb}
\usepackage{hyperref}
\usepackage{cite}
\usepackage{bbding}
\usepackage{makecell}

\begin{document}

\title[Microparticle charge measurements with quantum dots]{On the optical measurement of microparticle charge using quantum dots}

\author{M.Y. Pustylnik$^1$, Z. Marvi$^2$, J. Beckers$^2$}
\address{$^1$Institut f\"ur Materialphysik im Weltraum, Deutsches Zentrum f\"{u}r Luft- und Raumfahrt (DLR), M\"{u}nchener Stra\ss e 20, 82234 We\ss ling, Germany}
\address{$^2$Department of Applied Physics, Eindhoven University of Technology, P.O. Box 513, 5600MB Eindhoven, The Netherlands}
\ead{mikhail.pustylnik@dlr.de}
\vspace{10pt}
\begin{indented}
\item[]
\end{indented}

\begin{abstract}
We investigated the possibility of using a layer of quantum dots (QDs) deposited on the microparticle surface for the measurement of the charge the microparticle acquires when immersed into a plasma.
To that end, we performed the calculations of the Stark shift of the photoluminescence spectrum of QDs caused by the fluctuating local electric field.
In our calculations, we assumed the plasma-delivered surplus electrons to be distributed on the surface of a microparticle.
According to our calculations, the Stark shift will acquire measurable values when the lifetime of the quasi-stationary configuration of the surplus electrons will be determined by their diffusion along the surface.
Experiments with flat QD-covered floating plasma-facing surfaces suggest that measurable Stark shift of the photoluminescence spectrum can be achieved.
Based on our model, modern microscopic plasma-surface interaction theories and analysis of the experiments, we suggest the possible design of the charge microsensor, which will allow to measure the charge accumulated on its surface by means of visible-light optics.
\end{abstract}
\maketitle
\hypersetup{breaklinks=true}

%\maketitle
\section{Introduction}
\label{Sec:Intro}
Dusty plasmas are ionized gaseous media containing micro- or nanometer-sized solid particles.
They are ubiquitous.
They can be found in natural systems, such as interstellar clouds, planetary rings, noctilucent clouds, as well as in industrial plasma reactors \cite{Tsytovich1997, vladimirov2005physics}.
The charge of the dust particle is one of the most important characteristics of the dust component.
It enters into all dust-related quantities, e.g. electrostatic and ion drag forces \cite{Fortov2004rev}, coupling parameter \cite{Ikezi1986}, velocities of the wave modes \cite{Rao1990, Zhdanov2003}, absorption rates of plasma electrons and ions \cite{Gozadinos2003, Pustylnik2017}, etc.
Many phenomena in dusty plasmas are supposed to be the consequence of, e.g., charge fluctuations \cite{Vaulina1999}, charge gradients \cite{Vaulina2017} or delayed charging \cite{Nunomura1999, Pustylnik2006}.
\\ \indent
%**************************************************************************** 
In laboratory experiments, the charge is usually measured by means of dynamical methods. 
Dynamical methods can be either passive, as e.g. in \cite{Nosenko2018}, where the charge is determined from the spectrum of phonons spontaneously excited in a 2D monolayer of microparticles, or active, in which a response of the microparticles to an assumingely known disturbance is observed \cite{Trottenberg1995, Fortov2001, Fortov2004, Khrapak2005, Antonova2019}.
Such methods have obvious disadvantages.
Some methods rely on complicated models of microparticle dynamics (e.g., phonons in \cite{Nosenko2018}), some involve the interparticle interaction potential \cite{Fortov2004rev, Nosenko2018}. 
A certain group of methods requires assumptions on the spatial distribution of plasma parameters \cite{Trottenberg1995, Fortov2001} and/or rely on the exact values of plasma parameters \cite{Antonova2019}.
Values of the plasma parameters are often taken from measurements in dust-free plasmas, whereas it has been shown that dust strongly affects the local values of plasma parameters as well as their global distributions \cite{Mitic2009, Pustylnik2017, Pikalev2019}.
All the above-mentioned issues limit the accuracy of the measurements.
Also, in the dynamical methods, the measurement obviously occurs on the timescales of dust dynamics and therefore, variations of dust charge on plasma timescales can only be hypothesized, but not observed.
\\ \indent
%**************************************************************************** 
Importance of the dust charging problem found its reflection in the Consensus Study Report of the US National Academies of Sciences, Engineering and Medicine \cite{NAPPlasmaRep2020}.
The Committee on a Decadal Assessment of Plasma Science graded this problem as one of the future challenges.
The Report, in particular, states:
``While a number of techniques are used to make charge measurements (e.g., two-particle collisions and/or resonant oscillations), these measurements give the charge-to-mass ratio, and an assumed particle mass is then used to determine charge. 
A future challenge is the development of non-invasive, non-perturbative techniques that independently determine the particle mass and charge.'' 
\\ \indent
%**************************************************************************** 
Recently, significant theoretical efforts have been undertaken to explore the possibilities of optical detection of the dust charge.
The surplus electrons modify the dielectric permittivity of the dust particle material or surface conductivity of the microparticle and therefore affect some of the spectral features of the scattering.
This approach should allow to avoid all the limitations of the dynamical methods. 
In \cite{Heinisch2012, Heinisch2013}, excitonic resonance was considered, whereas \cite{Vladimirov2016, Vladimirov2017} investigated the possibility of usage of the surface plasmon resonance.
Experimental attempts are now limited to \cite{Krger2018} where charge related shift in the infrared absorption spectra could not be observed for $70-90$~nm size alumina dust particles.
\\ \indent
%**************************************************************************** 
In this paper, we propose a different view on the problem of optical measurement of the dust charge.
Instead of investigating the possibilities of the charge measurements on the dust particles consisting of uniform materials, we suggest to design a microparticle in such a way that it will work as an optical surface charge microsensor.
In principle, this approach has already been suggested in \cite{Thiessen2014}, where the usage of core-coat microparticles was suggested.
Our intention is to design a microsensor that would use spectral features in the visible range and not in the infrared as in \cite{Thiessen2014}.
This will allow to use the same illumination source for the observation of the microparticle dynamics and for the measurement of the microparticle charge.
\\ \indent
%**************************************************************************** 
Requirement of a special surface design will not allow us to address the direct needs of researchers working with natural dusty plasmas or particle growing plasmas.
However, development of such charge microsensors will allow to get experimental insights into many problems of dusty plasmas which up to now remain only in theoretical investigation.
The field of complex plasmas, in which microparticle suspensions in plasmas are used as particle-resolved models of condensed matter \cite{ILMRbook, FMbook}, will get a new experimental dimension since not only the microparticle positions and velocities, but also microparticle charges will become observable by optical means.
\\ \indent
%**************************************************************************** 
Such microsensors may be used in plasmas as small floating probes, which can be non-invasively placed into any region of the plasma by means of optical tweezers \cite{Schneider2018}.
Therefore, their development can potentially affect not only dusty plasma research, but the entire plasma diagnostics technology and through that the entire plasma science and plasma technology.
\\ \indent
%**************************************************************************** 
We propose here to modify the surface of the microparticles by coating them with a layer of quantum dots (QDs) \cite{Reed1988}.
QDs are semiconductor nanocrystalls whose electrical and optical properties depend on their size \cite{Ekimov1981, Gaponenko1998}.
The QDs exhibit photoluminescence properties:
Absorption of the incident light leads to the formation of electron-hole pairs, and their recombination produces monochromatic light, whose wavelength depends (among other parameters) on the QD size \cite{Yu2003}.
In the external electric field, quantum states of the hole and valence electrons are subject to the so-called quantum-confined Stark effect \cite{Ekimov1990}.
The photoluminescence spectrum will therefore experience Stark shift.
Due to much better stability, QDs are potential candidates for the substitution of traditional organic charge-sensitive dyes used for measurements of voltages in biological cells \cite{Efros2018}.
\\ \indent
%**************************************************************************** 
Very recently \cite{Zahra}, it was experimentally shown that the quantum-confined Stark effect of the photoluminescence of the QDs deposited on a plasma-facing surface can be used for the detection of microscopic electric fields.
Extending the approach of \cite{Zahra}, we suggest to use the QDs deposited on the surface of microparticles immersed in plasmas.
Using a simple model for the Stark shift of QD photolumeniscence spectrum, which takes into account fluctuations of the microscopic electric field, we propose the design of a microsensor that would exhibit measurable Stark shifts under typical plasma conditions.
\\ \indent
%**************************************************************************** 
Our paper is organized as follows: 
In \Sref{Sec:Mod}, we describe the model of the microscopic electric field on the microparticle surface.
In \Sref{Sec:QD}, we give our considerations regarding the QDs, which could be used for the coating of the microparticle surface.
In \Sref{Sec:RD}, we show the statistics of the calculated microscopic electric field, discuss the characteristic timescales of the changes in the electron configuration on the microparticle surface and show the calculated Stark shifts.
In \Sref{Sec:EC}, we consider different experimental aspects of utilizing QD-based microsensors in plasmas and propose the microsensor design.
\Sref{Sec:Conc} concludes the article.
\section{Model}
\label{Sec:Mod}
\subsection{Microscopic electric field}
\begin{figure}[t!]
\includegraphics[width=8.6cm]{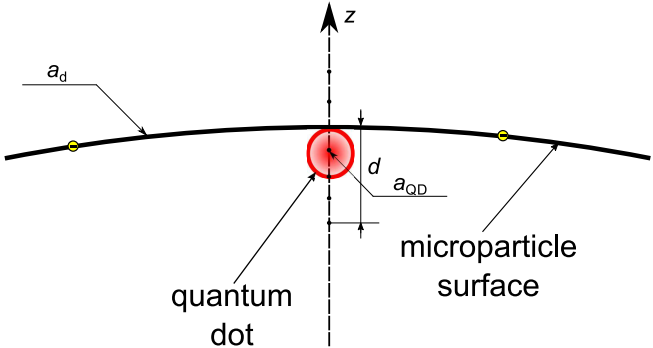}
\caption{
Scheme for the calculation of the electric field fluctuations on the surface of a spherical microparticle with the radius $a_{\rm d}$.
The electric field from every quasi-static electron configuration is calculated on the polar axis $z$ at a depth $h$ below the microparticle surface.
A quantum dot of radius $a_{\rm QD}$ is supposed to sit at the depth $h=a_{\rm QD}$.}
\label{Fig: Sch}
\end{figure}

Let us consider a microparticle of radius $a_{\rm d}$ immersed in a laboratory discharge plasma.
In such conditions, the microparticle normally acquires a negative charge of $Z_{\rm d}$ elementary charges.
This negative charge is represented by $Z_{\rm d}$ electrons.
Although, according to the modern theory \cite{PhysRevB.85.075323}, this is not always so, we suppose all these surplus electrons to be randomly distributed on the surface of a microparticle.
We will revisit the applicability of this assumption in \Sref{Sec:MD}.
\\ \indent
%**************************************************************************** 
The average macroscopic surface electric field associated with the microparticle charge is
\begin{equation}
E_{\rm d} = \frac{Z_{\rm d}e}{4\pi \epsilon_0 a_{\rm d}^2}.
\label{Eq:Ed}
\end{equation}
Debye shielding is neglected here for simplicity.
Average distance between the electrons sitting on the microparticle surface
\begin{equation}
l = \sqrt{\frac{4\pi a_{\rm d}^2}{Z_{\rm d}}}
\label{Eq:l}
\end{equation}
is for typical charges and radii of the order of several tens nm and is therefore much larger than the typical diameter of a QD ($2a_{\rm QD}$).
Therefore, the electric field sensed by a quantum dot (QD) placed close to the microparticle surface will be subject to large fluctuations.
We developed a calculation procedure to collect the statistics of those fluctuations.
\\ \indent
%****************************************************************************
In this procedure, $Z_{\rm d}$ point-like electrons were first randomly distributed over a sphere of radius $a_{\rm d}$. 
The obtained electron configuration was considered as quasi-stationary.
Then, we calculated the electric field on the polar axis $z$ at the depth $d$ below the microparticle surface (\Fref{Fig: Sch}).
Cartesian components of the electric field $E_{\rm x}$, $E_{\rm y}$ and $E_{\rm z}$ at this position are calculated in the following way:
\begin{eqnarray}
E_{\rm x(y)} = -\frac{e}{4\pi \epsilon_0}\sum^{Z_{\rm d}}_{i=1}\frac{x(y)_{\rm i}}{\left(x_{\rm i}^2+y_{\rm i}^2+\left(z_{\rm i}-a_{\rm d}+d \right)^2\right)^{\frac{3}{2}}},\\
%_{\rm y} = -\frac{e}{4\pi \epsilon_0}\sum^{Z_{\rm d}}_{i=1}\frac{y_{\rm i}}{\left(x_{\rm i}^2+y_{\rm i}^2+\left(z_{\rm i}-a_{\rm d}+d \right)^2\right)^{\frac{3}{2}}},\\
E_{\rm z} = \frac{e}{4\pi \epsilon_0}\sum^{Z_{\rm d}}_{i=1}\frac{a_{\rm d}-d-z_{\rm i}}{\left(x_{\rm i}^2+y_{\rm i}^2+\left(z_{\rm i}-a_{\rm d}+d \right)^2\right)^{\frac{3}{2}}},
\end{eqnarray}
where $x_{\rm i}$, $y_{\rm i}$ and $z_{\rm i}$ are the Cartesian coordinates of the electrons on the sphere ($x_{\rm i}^2+y_{\rm i}^2+z_{\rm i}^2=a_{\rm d}^2$).
Dielectric constant of the microparticle material is supposed here to equal unity. 
The absolute value of the electric field is then
\begin{equation}
E = \sqrt{E_{\rm x}^2+E_{\rm y}^2+E_{\rm z}^2}.
\end{equation}
Repeating the calculation for many random configurations allows to collect the statistics of electric field.
\\ \indent
%****************************************************************************
\subsection{Stark effect}
Now, let us consider a spherical QD whose center is located below the microparticle surface at $d=a_{\rm QD}$ (\Fref{Fig: Sch}).
For a certain static electron configuration, this QD will sense the electric field $E\left(d=a_{\rm QD}\right)$.
Under the effect of the aforementioned electric field, the QD photoluminescence wavelength will undergo a red shift due to the quantum-confined Stark effect \cite{Efros2018}.
This shift is expressed as
\begin{equation}
\Delta \lambda = 0.03\frac{\lambda^2}{hc}\left(m_{\rm v}+m_{\rm h}\right)a_{\rm QD}^4\left(\frac{2\pi eE}{h}\right)^2,
\label{Eq:Stark}
\end{equation}
where $m_{\rm v}$ and $m_{\rm h}$ are the effective masses of a valence electron and a hole, respectively and $\lambda$ is the photoluminescence wavelength.
In \sref{Sec:Stark}, we will investigate if this Stark shift could be used to measure the charge on microparticles immersed in laboratory plasmas.
\\ \indent
%****************************************************************************
We note that in an experiment, $\Delta \lambda$ will be measured over long ($>100$~ms \cite{Zahra}) exposure times.
It will be therefore integrated over many quasi-static electron configurations as well as over many radiation acts.
The result of averaging will depend on the relation between the radiative lifetime of a QD excited level $t_{\rm rad}$ and the lifetime of a quasi-static electron configuration on the microparticle surface $t_{\rm econf}$.
This issue will be discussed in \sref{Sec:tscales}.
\\ \indent
%****************************************************************************
\section{Quantum dots}
\label{Sec:QD}
The QD size $a_{\rm QD}$ does not affect the electric field fluctuations. 
On the other hand, as follows from \Eref{Eq:Stark}, the Stark shift is proportional to  $a_{\rm QD}^4$.
Therefore, to maximize the effect, we choose the largest commercially available core-shell CdSe/ZnS QDs with $a_{\rm QD}=3.3$~nm and $\lambda=650$~nm \cite{Yu2003, plasmachem}.
With the typical full width at half maximum of $30$~nm, the photoluminescence band will not overlap with the spectral lines corresponding to 2p$\to$1s transitions of argon, which are typically bright in the conditions of complex plasma experiments.
In the complex plasma experiments, argon is one of the most frequently used plasma-forming gases.
Valence electron and hole masses for CdSe are $m_{\rm v}=0.13 m_{\rm e}$ and $m_{\rm h}=0.45 m_{\rm e}$, respectively \cite{effmasses}.
QD parameters given above will be used for the calculations of the Stark shift throughout the paper.
%**************************************************************************** 
\section{Results and discussion}
\label{Sec:RD}
\subsection{Fluctuations of the electric field}
\label{Sec:Fluc}
\begin{figure}[t!]
\includegraphics[width=8cm]{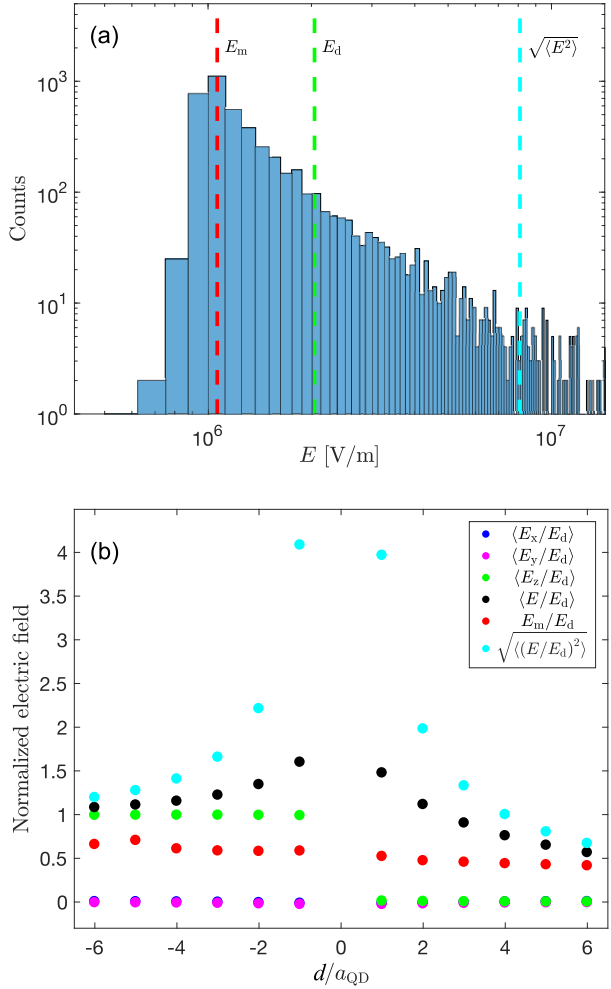}
\caption{
Results of electric field calculations for $a_{\rm d}=4.6$~$\mu$m and $Z_{\rm d}=3\times10^4$:
(a) Histogram of electric field $E$ at the location of a QD ($d=a_{\rm QD}$).
A maximum is observed at $E=E_{\rm m}$.
Values of $E_{\rm m}$, $E_{\rm d}$ and rms electric field averaged over $5000$ quasi-static electron configurations $\sqrt{\langle E^2\rangle}$ are shown by red, green and blue vertical dashed lines, respectively.
At $E\gtrsim E_{\rm m}$, the histogram exhibits almost power law decay.
(b) Calculated normalized values of the electric field averaged over $5000$ quasi-static electron configurations at different axial position.
In the vicinity of the microparticle surface, $\langle E \rangle$ and exceeds the average value of axial electric field $E_{\rm d}$, whereas the most probable electric field $E_{\rm m}$ lies below $E_{\rm d}$.
The rms electric field exceeds $E_{\rm d}$ several times.}
\label{Fig:efVH}
\end{figure}

To demonstrate the calculation of the electric field fluctuations, we used experimental data from \cite{Nosenko2018}, where the charge of melamineformaldehyde microspheres in a monolayer plasma crystal was measured by fitting the power spectrum of spontaneously excited phonons.
This is a very reliable dynamical technique based on the assumption of Yukawa interaction between the microparticles and not requiring the knowledge of plasma parameters.
For this example, we took $a_{\rm d}=4.6$~$\mu$m, $Z_{\rm d}=3\times10^4$ and generated $N_{\rm econf}=5\times10^3$~electron configurations.
The results are shown in \Fref{Fig:efVH} and summarized in \Tref{Tab:E}.
\Tref{Tab:E} contains also one more dataset based on recent measurements conducted in a microgravity complex plasma facility PK-4 on board the International Space Station \cite{Antonova2019}.
\\ \indent
%****************************************************************************
In \Fref{Fig:efVH}(a), we show the histogram of $E$ at the location of a QD.
At small electric fields, the histogram rapidly increases and reaches a maximum at the value of $E_{\rm m}\approx \frac{1}{2}E_{\rm d}$, which represents the most probable electric field among $N_{\rm econf}$ electron configurations.
At $E\gtrsim E_{\rm m}$, the histogram exhibits almost a power law decay.
\\ \indent
%****************************************************************************
In \Fref{Fig:efVH}(b), we show the quantities averaged over $N_{\rm econf}$ electron configurations: For any quantity $\mathcal{E}$,
\begin{equation}
\langle \mathcal{E}\rangle=\frac{1}{N_{\rm econf}}\sum_{j=1}^{N_{\rm econf}}\mathcal{E_{\rm j}},
\end{equation}
where $\mathcal{E_{\rm j}}$ is the value of this quantity for $j$-th electron configuration.
As expected, the transverse components $\langle E_{\rm x} \rangle$ and $\langle E_{\rm y}\rangle$ vanish below as well as above the microparticle surface, whereas the axial component  $\langle E_{\rm z}\rangle$ vanishes only below the microparticle surface and acquires the value of $E_{\rm d}$ above.
The absolute value of the electric field $\langle E \rangle$ exceeds $E_{\rm d}$ in the vicinity of the surface, tends to $E_{\rm d}$ above the microparticle surface and tends to vanish below the microparticle surface.
However, at the depth of $6a_{\rm QD}$, electric field of $\approx \frac{1}{2}E_{\rm d}$ can still be sensed.
The vanishing depth is obviously of the order of $l\approx 28a_{\rm QD}$ in this particular case (see \Tref{Tab:E}).
The most probable electric field $E_{\rm m}$ varies between $0.4E_{\rm d}$ and $0.7E_{\rm d}$ in the vicinity of the microparticle surface.
\\ \indent
%****************************************************************************
\begin{figure*}[h]
\includegraphics[width=14cm]{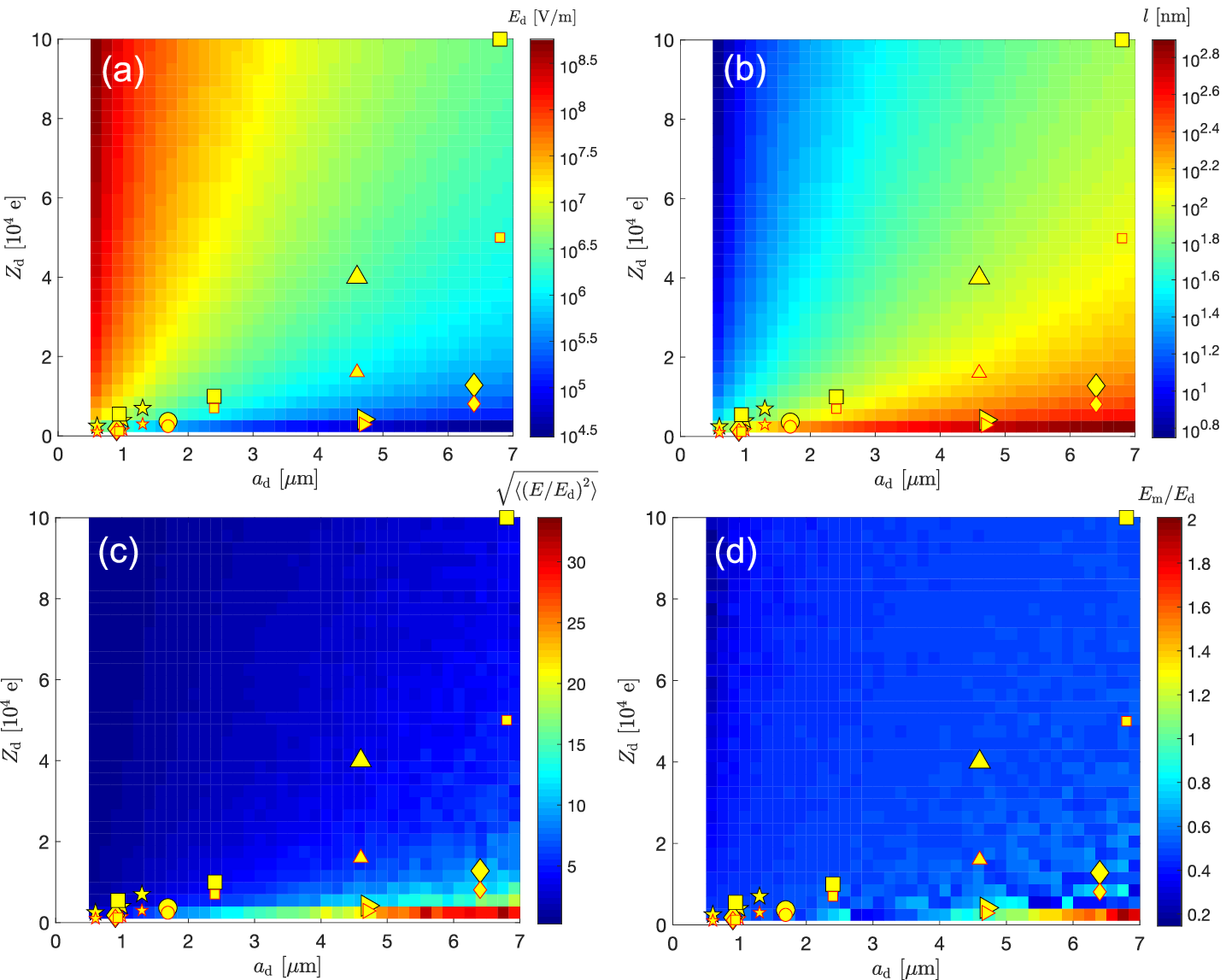}
\caption{Dependence of (a) macroscopic electric field on the microparticle surface $E_{\rm d}$ (\Eref{Eq:Ed}) and (b) mean distance between the electrons on the microparticle surface $l$ (\Eref{Eq:l}) on microparticle radius $a_{\rm d}$ and charge $Z_{\rm d}$.
Results of the calculations of (c) normalized local rms electric field sensed by a QD $\sqrt{\langle (E/E_{\rm d})^2\rangle}$ averaged over the quasi-static electron configurations and (d) normalized most probable electric field $E_{\rm m}/E_{\rm d}$ for different $a_{\rm d}$ and $Z_{\rm d}$.
Since both $\langle E/E_{\rm d}\rangle$ and $E_{\rm m}/E_{\rm d}$ are connected with the fluctuations of the local electric field, their dependence on $a_{\rm d}$ and $Z_{\rm d}$ is qualitatively similar to that of $l$.
Available literature data on microparticle charge in complex plasmas (\Tref{Tab:ChLit}) are plotted on all four plates.
Bigger symbols with black borderline and smaller symbols with red borderline correspond to the maximal and minimal $Z_{\rm d}$ values, respectively.}
\label{Fig:Zdad}
\end{figure*}
We have also collected the statistics of electric field fluctuations in the wide range of $Z_{\rm d}$ (between $10^3$ and $10^5$ elementary charges) and $a_{\rm d}$ (between $0.5$ and $7$~$\mu$m).
For each value of $Z_{\rm d}$, we selected $N_{\rm econf}$ in such a way that the product $N_{\rm econf}Z_{\rm d}=1.5\times10^8$.
\\ \indent
%****************************************************************************
\Fref{Fig:Zdad}(a) and (b) show the values of $E_{\rm d}$ and $l$ calculated with \Eref{Eq:Ed} and~\eref{Eq:l}, respectively.
Dependencies of $E_{\rm d}$ and $l$ on $a_{\rm d}$ and $Z_{\rm d}$ are trivial: $E_{\rm d}$ increases, and $l$ decreases with the increase of $Z_{\rm d}$ and decrease of $a_{\rm d}$.
\\ \indent
%****************************************************************************
The normalized rms electric field $\sqrt{\langle(E/E_{\rm d})^2\rangle}$ varies similarly to $l$ in the plane $\left(a_{\rm d}\right.$,~$\left.Z_{\rm d}\right)$, which is expected since the fluctuations increase with the increase of the mean distance between electrons on the microparticle surface.
We note that $\sqrt{\langle(E/E_{\rm d})^2\rangle}$ considerably exceeds unity in a wide range of $a_{\rm d}$ and $Z_{\rm d}$.
In the area of small $l$, $\sqrt{\langle(E/E_{\rm d})^2\rangle}$ approaches unity.
The smallest value of $l$ in our parameter space is $\approx 6$~nm, which is comparable to $a_{\rm QD}$.
Under such conditions, fluctuations of the electric field will be significantly suppressed already on the dimensions of a QD.   
The normalized most probable electric field $E_{\rm m}/E_{\rm d}$ exhibits a similar trend as $\sqrt{\langle(E/E_{\rm d})^2\rangle}$, but is considerably smaller in magnitude.
\\ \indent
%****************************************************************************
We have also collected the data on microparticle charges available in the literature (see \Tref{Tab:ChLit}).
\begin{table}
\begin{tabular}{||c|c|c|c||}
\hline
$a_{\rm d}$ [$\mu$m]	&\makecell{Minimal \\ $Z_{\rm d}$ [$10^4$e]}	&\makecell{Maximal \\ $Z_{\rm d}$ [$10^4$e]}	&\makecell{Symbol / \\ Reference}		\\ \hline
4.7					&0.29								&0.42								&$\rhd$ / \cite{Trottenberg1995}		\\ \hline
0.94 					&0.12								&0.55								&$\square$ / \cite{Fortov2001}			\\ \hline
2.4 					&0.7									&1.0									&$\square	$ / \cite{Fortov2001}			\\ \hline
6.8 					&5.0	 								&10.0								&$\square	$ / \cite{Fortov2001}			\\ \hline
0.9 					&0.115 								&0.18 								&$\Diamond$ / \cite{Fortov2004}		\\ \hline
6.4 					&0.81 								&1.28 								&$\Diamond$ / \cite{Fortov2004}		\\ \hline
0.6 					&0.1	 								&0.25 								&\FiveStarOpen / \cite{Khrapak2005}	\\ \hline
1.0 					&0.15 								&0.4	 								&\FiveStarOpen / \cite{Khrapak2005}	\\ \hline
1.3 					&0.3	 								&0.7	 								&\FiveStarOpen / \cite{Khrapak2005}	\\ \hline
4.6 					&1.6	 								&4.0	 								&$\triangle$ / \cite{Nosenko2018}		\\ \hline
1.7 					&0.23 								&0.35								&$\bigcirc$ / \cite{Antonova2019}		\\ \hline
\end{tabular}	
\caption{Data on the microparticle charges available from literature.
For each value of microparticle radius $a_{\rm d}$, a range of charges between minimal and maximal $Z_{\rm d}$ is defined.
These data are plotted in \Fref{Fig:Zdad}.}
\label{Tab:ChLit}
\end{table}
Since here, we are not interested in the physics of charging, but only in the value of $Z_{\rm d}$, we have collected papers, where $Z_{\rm d}$ was experimentally measured in a certain discharge parameter range (and in \cite{Antonova2019} even in different gases) so that a range of $Z_{\rm d}$ values for each value of $a_{\rm d}$ could be defined.
In~\cite{Trottenberg1995, Nosenko2018}, the experiments were performed in a capacitively coupled rf discharge, whereas Refs.~\cite{Fortov2001, Fortov2004, Khrapak2005, Antonova2019} deal with dc discharge.
In~\cite{Trottenberg1995, Nosenko2018}, the measurements were performed in the sheath region of the plasma, whereas
In~\cite{Antonova2019}, the microparticles were immersed in the bulk of the discharge under the microgravity conditions.
In~\cite{Fortov2001, Fortov2004}, the experiments were carried out in the standing striations of a dc discharge.
We plotted the data in all the four plates of \Fref{Fig:Zdad}.
\\ \indent
%****************************************************************************
\subsection{Characteristic timescales}
\label{Sec:tscales}
As mentioned above, the resulting Stark shift of the QD photoluminescence spectrum will depend on the relation between the radiation lifetime of a QD excited level $t_{\rm rad}$ and the lifetime of a quasi-stationary electron configuration $t_{\rm econf}$. 
The former time is quite well known for each particular type of QD and is usually $\sim10^{-8}$~s \cite{Efros2018}.
Obviously, $t_{\rm econf}$ will be determined by the fastest of two processes: plasma charging by electron and ion fluxes and diffusion of electrons along the microparticle surface.
\\ \indent
%****************************************************************************
The first process is more or less well studied.
The electric field sensed by a QD is mainly determined by the electrons in its vicinity. 
Therefore, it will significantly change only when those particular electrons recombine with impinging ions from the plasma.
This time can be expressed as \cite{Fortov2004rev}
\begin{equation}
t_{\rm ch}=\frac{Z_{\rm d}}{a_{\rm d}^2 n_{\rm e}\sqrt{\frac{8\pi kT_{\rm e}}{m_{\rm e}}}\exp{\left(-\frac{Z_{\rm d}e}{4\pi\epsilon_0 a_{\rm d}kT_{\rm e}}\right)}} ,
\end{equation}
where $n_{\rm e}$ and $T_{\rm e}$ are the plasma electron density and temperature, respectively.
This charging time usually varies from fractions of a $\mu$s to ms (see data in \Tref{Tab:E}).
To decrease it further, one will have to increase the plasma density several orders of magnitude which is usually not compatible with trapping the microparticles inside the plasma.
\\ \indent
%****************************************************************************
Diffusion of the electrons along the surface will be limited by the surface roughness and, therefore, depends on the particular type of microparticles.
As an example, we take the natural surface roughness of the melamineformaldehyde microparticles which was studied in \cite{Semenov2018} using atomic force microscopy (AFM) and in \cite{Karasev2016, Karasev2017} using the scanning electron microscopy (SEM).
According to the AFM images, the surface is formed by globules with the typical size $s_{\rm g}\sim40$~nm, which create average surface roughness $s_{\rm r}\approx2$~nm.
SEM gives larger roughness $s_{\rm r}\approx10$~nm and similar globule size.
The potential barriers associated with this roughness will be of the order of $eE_{\rm d}s_{\rm r}\sim10^{-3}$~eV. 
Therefore, electrons that are fully thermalized with the surface \cite{PhysRevB.85.075323} will be able to freely diffuse along the surface and will not be trapped in the roughness.
The diffusion-limited lifetime of a quasi-steady-state electron configuration $t_{\rm diff}$ can be estimated as the time required for an electron to diffuse over a distance $l$:
\begin{equation}
\label{Eq:tdiff}
t_{\rm diff}=\frac{l^2}{s_{\rm g}v_{\rm eth}},
\end{equation}
where $v_{\rm eth}=\sqrt{\pi k_{\rm B}T/2m_{\rm e}}$ is the two-dimensional thermal velocity of electrons and $s_{\rm g}$ plays the role of the mean free path.
For typical $l\sim100$~nm and $T=300$~K, $t_{\rm diff}\approx3\times10^{-12}$~s, which is far below $t_{\rm rad}$.
\\ \indent
%****************************************************************************
As the estimations suggest that $t_{\rm rad}\ll t_{\rm econf}$, in absence of direct experimental measurement of $t_{\rm econf}$, we, in principle, have to consider an arbitrary relation between  $t_{\rm rad}$ and $t_{\rm diff}$.
However, in this case, the calculation of the Stark shift of a QD photoluminescence spectrum requires the treatment of the surplus electron kinetics in its entire complexity and represents quite a difficult task.
Therefore, we will concentrate our attention on two asymptotic regimes: ``radiation-dominated'', in which $t_{\rm rad}\ll t_{\rm econf}$, and ``diffusion-dominated'', in which $t_{\rm econf}\ll t_{\rm rad}$.
%****************************************************************************
\subsection{Stark shift of quantum dots}
\label{Sec:Stark}
\begin{figure}[t!]
\includegraphics[width=7.6cm]{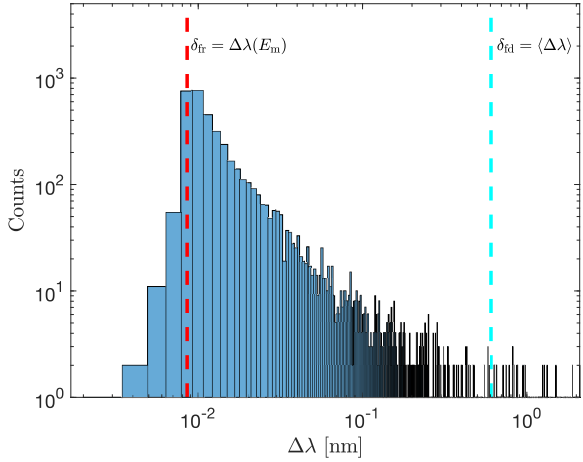}
\caption{Histogram of QD photoluminescence Stark shifts for different quasi-stationary electron configurations at $a_{\rm d}=4.6$~$\mu$m and $Z_{\rm d}=3\times10^4$ calculated using \Eref{Eq:Stark} from the electric field histogram in \Fref{Fig:efVH}(a).
Values $\delta_{\rm fr}$ and $\delta_{\rm fd}$ correspond to the Stark shifts in ``radiation-dominated'' and ``diffusion-dominated'' regimes, respectively.
}
\label{Fig:stHist}
\end{figure}

In the radiation-dominated regime, each radiation act will take place at a certain quasi-stationary electron configuration.
Therefore, a time-averaged photoluminescence spectrum will only reflect the statistics over quasi-stationary electron configurations and the observed spectral shift $\delta_{\rm fr}=\Delta\lambda\left(E_{\rm m}\right)$ will correspond to the most probable local electric field. 
The histogram of the Stark shifts in \Fref{Fig:stHist} calculated using \Eref{Eq:Stark} for the selected example of $a_{\rm d}=4.6$~$\mu$m and $Z_{\rm d}
=3\times10^4$ represents nothing but a photoluminescence spectrum of a QD on the surface of this microparticle in the radiation-dominated regime and provided the QD photoluminescence is monochromatic at zero electric field.
In the diffusion-dominated regime, all possible quasi-stationary electron configurations are realized within the lifetime of an excited state and the photoluminescence spectral line will shift with the value of $\delta_{\rm fd}=\langle\Delta\lambda\rangle\sim \langle E^2\rangle$.
\\ \indent
%****************************************************************************
\begin{figure*}
\includegraphics[width=16cm]{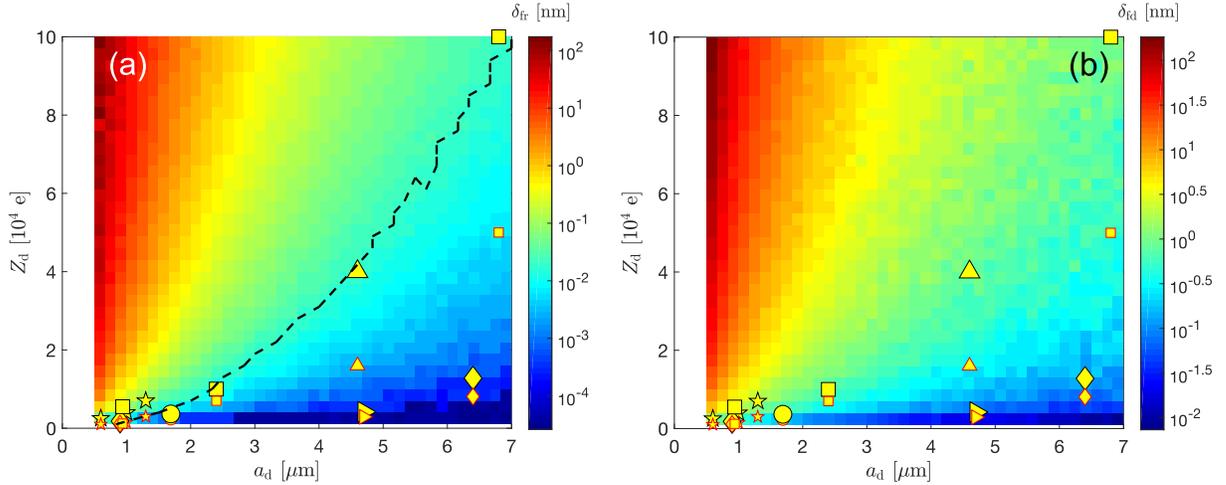}
\caption{Values of the Stark shift on a QD with the radius $a_{\rm QD}=3.3$~nm on the surface of a microparticle in the (a) radiation-dominated and (b) diffusion-dominated regimes. 
Dashed line in plate (a) represents the experimentally estimated detection limit of $0.02$~nm \cite{Zahra}.
In the area above the dashed lines, the Stark shift can be measured. 
In the area below the dashed lines, the Stark shift is obscured by noise.
In plate (b), the Stark shift can be measured in the entire parameter range.
The data from \Tref{Tab:ChLit} are also plotted.
Bigger symbols with black borderline and smaller symbols with red borderline correspond to the maximal and minimal $Z_{\rm d}$ values, respectively.}
\label{Fig:deltas}
\end{figure*}
We have calculated $\delta_{\rm fr}$ and $\delta_{\rm fd}$ in the wide range of $a_{\rm d}$ and $Z_{\rm d}$.
The results are shown in \Fref{Fig:deltas}.
Almost for the entire parameter range, $\delta_{\rm fr}<\delta_{\rm fd}$ and therefore $\delta_{\rm fr}$ and $\delta_{\rm fd}$ can be understood as lower and upper estimations of the Stark shift, respectively.
\\ \indent
%****************************************************************************
\section{Experimental considerations}
\label{Sec:EC}
\subsection{Detection limit of the Stark shift}
\label{Sec:DL}
\begin{table}
\begin{tabular}{||l|c|c||}
\hline
{\bf Parameter}					&{\bf Data of \cite{Nosenko2018}}	&{\bf Data of \cite{Antonova2019}}	\\ \hline
$n_{\rm e}$ [m$^{-3}$]			&$1.0\times10^{15}$				&$4.0\times10^{14}$				\\ \hline
$T_{\rm e}$ [eV]				&$1.3$ 						&$4.5$						\\ \hline
$p$ [Pa]						&$0.4$ 						&$40$						\\ \hline
$a_{\rm d}$ [$\mu$m]			&$4.6$ 						&$1.7$						\\ \hline
$Z_{\rm d}$ [e]					&$30000$						&$2000$						\\ \hline
$l$ [nm]						&$94$						&$132$						\\ \hline
$E_{\rm d}$ [V/m]				&$2.0\times10^6$				&$1.0\times10^6$				\\ \hline
$E_{\rm m}$ [V/m]				&$1.0\times10^6$				&$5.6\times10^5$				\\ \hline
$\langle E\rangle$ [V/m]			&$2.9\times10^6$				&$1.7\times10^6$				\\ \hline
$\sqrt{\langle E^2\rangle}$ [V/m]	&$8.1\times10^6$				&$5.8\times10^6$				\\ \hline
$t_{\rm ch}$ [s]					&$8.1\times10^{-4}$				&$5.9\times10^{-7}$				\\ \hline
$t_{\rm heat}$ [s]				&$0.43$						&$0.031$						\\ \hline
$\delta_{\rm fr}$ [nm]				&$0.010$						&$0.0023$					\\ \hline	
$\delta_{\rm fd}$ [nm]			&$0.61$						&$0.31$						\\ \hline
\end{tabular}	
\caption{Examples of experimental conditions under which the microparticle charges were measured by dynamical methods. Parameters relevant for the QD Stark shift measurement were calculated for those conditions with $a_{QD} = 3.3$~nm and $\lambda = 650$~nm. For the calculations of $t_{\rm heat}$, the microparticle material was assumed to be melamineformaldehyde with density $\rho=1500$~kg/m$^3$ \cite{microparticles} and specific heat $C=1500$~J/kg$\times$K \cite{designerdata}, gas was assumed to be argon with the atomic mass $M=40$~a.m.u. and temperature $T_{\rm g}=300$~K.}
\label{Tab:E}
\end{table}
To be able to judge whether the calculated Stark shift can be detected in an experiment, we estimate the detection limit based on the results of \cite{Zahra}.
In that experiment, time-resolved measurements of the photoluminescence spectrum of QDs on a substrate surface exposed to a capacitively coupled RF plasma were performed. 
CdSe/ZnS quantum dots ($a_{\rm QD}=1.3$~nm) were excited using a $405$~nm pulsed laser, after which temporal evolution of the photoluminescence spectra have been measured using a monochromator backed by an ICCD camera.
The spectral resolution of that system was $0.1$~nm.
The spectral axis of the ICCD camera had a pixel resolution of $0.05$~nm/pixel.
\\ \indent
%****************************************************************************
Indeed, a red shift of the photoluminescence spectrum was observed during plasma exposure, while before the plasma ignition, the photoluminescence spectrum stayed unchanged. 
From Figure 4 in \cite{Zahra}, it can be seen that the noise level in the measured time series of photoluminescence spectrum peak position is about $0.01$~nm. 
Therefore, we can accept $0.02$~nm as a realistic detection limit for the Stark shift.
\\ \indent
%****************************************************************************
In both plates of \Fref{Fig:deltas}, we have plotted the points corresponding to the typical value of the microparticle charges in complex plasma experiments (\Tref{Tab:ChLit}).
As seen from \Fref{Fig:deltas}(a), $\delta_{\rm fr}$ for most of the experimental point appears to be unmeasurable.
On the other hand, as shown in \Fref{Fig:deltas}(b), if the experimental points are placed into the fast diffusion regime, the Stark shift will become measurable in the entire parameter range.
The same is demonstrated in \Tref{Tab:E} for the two particular examples of the data from \cite{Nosenko2018, Antonova2019}.
%****************************************************************************
\subsection{Heating of a microparticle surface in a plasma} 
It was shown in experiments \cite{Swinkels2000, Maurer2008} and theoretical works \cite{Khrapak2006} that microparticles immersed in plasmas should heat up several tens K with respect to the neutral gas due to the kinetic and potential energy, the electrons, ions and excited atoms deposit on their surfaces.
On the other hand, QD photoluminescence is known to experience the red shift on the increase of the QD temperature \cite{Walker2003, Valerini2005, Albahrani2018, Zahra}.
The magnitude of the shift for CdSe/ZnS QDs is $\approx0.1$~nm/K \cite{Zahra}.
Therefore, the Stark shift due to the microparticle charge will be significantly smaller than the thermal shift.
\\ \indent
%****************************************************************************
As it was shown in \cite{Zahra}, after switching on the plasma, the peak wavelength of the photoluminescence spectrum of the plasma-facing QDs varies on two very different timescales:
First, it exhibits an immediate increase associated with the quatum-confined Stark effect and then goes into the long-term growing trend determined by heating of the substrate.
Photoluminescence of the QDs deposited on the surface of microparticles immersed in plasmas should be a subject to similar behavior, which makes the problem of distinguishing between the charge- and heat-induced red shift not so severe.
\\ \indent
%****************************************************************************
The microparticles are only cooled by the surrounding neutral monoatomic gas and by radiation.
Let us assume that $\left|T_{\rm g}-T_{\rm d}\right|\ll T_{\rm g}$, where $T_{\rm g}$ and $T_{\rm d}$ are the gas and microparticle surface temperatures, respectively.
Then, linearizing the radiative flux from the microparticle surface $\sigma\left(T_{\rm d}^4-T_{\rm g}^4\right)$, we express the energy balance on the surface: $Cm_{\rm d}\partial T_{\rm d}/\partial t=Q_{\rm pl}+2a_{\rm d}^2\left(p\sqrt{8\pi k_{\rm B}/MT_{\rm g}}+8\pi \sigma T_{\rm g}^3\right)\left(T_{\rm g}-T_{\rm d}\right)$, where $C$ is the specific heat of the microparticle material, $m_{\rm d}$ is the microparticle mass, $M$ is the mass of a gas atom and $Q_{\rm pl}$ is the (constant in time) thermal flux arriving at the microparticle surface from the plasma.
Then the characteristic heating time will be
\begin{equation}
t_{\rm heat} = \frac{C\rho a_{\rm d}}{3}\frac{1}{p\sqrt{2k_{\rm B}/\pi MT_{\rm g}}+4\sigma T_{\rm g}^3},
\end{equation}
where $\rho$ is the density of the microparticle material.
\\ \indent
%****************************************************************************
Calculations for the two representative examples from \cite{Nosenko2018, Antonova2019} are given in \Tref{Tab:E}.
Obviously, $t_{\rm heat}$ is several orders of magnitude larger than $t_{\rm ch}$.
Therefore, by pulsing the discharge with the frequency between $t_{\rm heat}^{-1}$ and $t_{\rm ch}^{-1}$, it will be possible to distinguish between the thermal red shift and charge-induced Stark shift.
%****************************************************************************
\subsection{Microsensor design}
\label{Sec:MD}
\begin{figure}
\includegraphics[width=5cm]{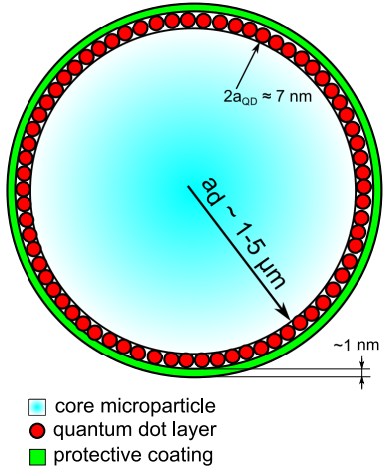}
\caption{Schematically represented proposed design of a surface charge microsensor in which the surplus electrons will be distributed in the vicinity of the surface and their arrangement will closely correspond to the surface arrangement of the surplus electrons in the proposed model. 
A microparticle (of $1-5$~$\mu$m radius) is coated by a layer of semiconductor QDs (of $6.6$~nm diameter). 
The layer of QDs is then protected against plasma damage and electron electron penetration by a thin ($\sim 1$~nm) layer of material with negative electron surface affinity.}
\label{Fig:musensor}
\end{figure}

We have seen in \Sref{Sec:DL} that the Stark shift due to the surface charge on a microparticle can in principle be measurable.
Moreover, estimations of the surplus electron diffusion time on the surface of a microparticle suggest that the QDs will operate in the diffusion-dominated regime resulting in a large measurable Stark shift.
However, we mentioned in \Sref{Sec:Mod} that according to \cite{PhysRevB.85.075323}, the assumption of the surface surplus charge is not always correct.
Indeed, the result of the theoretical calculation in \cite{PhysRevB.85.075323} is that only for the surfaces with negative electron affinity the surplus charge will accumulate in a thin layer fractions of a nm above the crystallographic surface.
For the surfaces with positive electron affinity, the surplus charge penetrates to millimeter depths into the plasma-facing material.
For micrometer-sized microsensors, that would mean almost homogeneous distribution of the surplus electrons in its volume, which will drastically reduce the fluctuations of the electric field and therefore decrease chances to detect the Stark shift. 
\\ \indent
%****************************************************************************
On the other hand, the outcome of the experiments in \cite{Zahra}, where the QDs were deposited on a silicon substrate, at a first glance contradicts the statement above.
Silicon has an electron affinity of $4.05$~eV \cite{ioffeSi} and, therefore, the surplus electrons should be distributed in rather a thick substrate layer.
Nevertheless, the measured charge-induced Stark shift even exceeded $\delta_{\rm fd}$ estimated using the surface-charge assumption about a factor of two.
Since the QD-coated surface was floating in the plasma, its potential could not be measured independently.
However, the fact that in that work $\delta_{\rm fd}$ exceeds $\delta_{\rm fr}$ more than two orders of magnitude suggests that the surplus electrons really concentrate themselves in the vicinity of the surface.
\\ \indent
%****************************************************************************
A possible explanation of this contradiction can be found if one takes into account large electron affinity of the QD material, CdSe, which is $4.95$~eV \cite{Swank1967} and therefore larger than that of the silicon substrate.
It could be that the potential barrier between the QDs and the substrate keeps most of the surplus electrons in the vicinity of the sensing QDs.
Excessive (compared to the estimated $\delta_{\rm fd}$) measured Stark shift could be a consequence of the underestimated surface potential, but could also point to the incorrect account for the electron configurations in which the surplus electrons closely approach the QDs.
Estimations in \cite{Zahra} show that correct quantum-mechanical treatment of such configurations may result in the reduction of the discrepancy between the observed and the estimated Stark shift.
\\ \indent
%****************************************************************************
Another problem of QD-sensing in plasmas demonstrated in \cite{Zahra} is the irreversible damage the plasma causes on the sensing QDs.
Small, but spectrally detectable damage was observed already at $30$~s exposure of the substrate to the plasma.
To mitigate this unavoidable problem, protective coatings deposited on top of the QD layers can be used.
The electron surface affinity of the coating material (as well as that of the microparticle material) should stay below the electron affinity of the QD material.
\\ \indent
%****************************************************************************
In a special case when the thin ($\sim1$~nm thickness) protective coating is made of a material with the negative electron surface affinity (\Fref{Fig:musensor}), the surplus electrons will be kept above the QD layer and their configuration will closely approach the surplus charge configuration of the present model. 
Appropriate coating materials will then be diamond, boron nitride or alkaline earth oxides \cite{PhysRevB.85.075323}.
The electron surface affinity of the microparticle material should not than play a significant role.
\\ \indent
%****************************************************************************
\section{Conclusion}
We investigated the possibility of the measurement of charges of microparticles immersed in a plasma by means of QDs deposited on their surface.
For that, we developed a simple model for the calculation of the Stark shift of the photoluminescence spectrum of QDs located in the vicinity of the surface of a charged micron-sized particle.
Locations of the surplus electrons were restricted to the microparticle surface.
The model takes into account fluctuations of the local electric field, which were considered by acquiring the statistics over randomly generated quasi-stationary electron configurations.
\\ \indent
%****************************************************************************
For the calculations of the Stark shift, we considered two asymptotic regimes.
In one of them, which we designated as radiation-dominated, the radiation lifetime of an excited state of a QD (being of the order of $10^{-8}$~s) is much smaller than the lifetime of a quasi-stationary electron configuration.
In the other regime, which we designated as diffusion-dominated, on the contrary, the lifetime of a quasi-stationary electron configuration is much smaller than the radiation lifetime of QD's excited state.
  \\ \indent
%**************************************************************************** 
 There are two processes, which determine the lifetime of a quasi-stationary electron configuration on the surface of a microparticle: charging by plasma electrons and ions and diffusion of the microparticles along the surface.
According to our estimations, the former appears to be much slower ($10^{-7}-10^{-3}$~s) than the radiative decay of QD's excited state, whereas the latter appears to be much faster ($\sim10^{-12}$~s) than the radiative decay of QD's excited state.
 \\ \indent
%**************************************************************************** 
In the diffusion-dominated regime, the Stark shift appears to be significantly larger than in the radiation-dominated regime.
We also found, that in many experimental cases known from literature, for which the charge was measured by dynamical methods, the QDs deposited on the microparticle surface would exhibit a measurable Stark shift of photoluminescence only in the diffusion-dominated regime, whereas in the radiation-dominated regime, the shift will be in most cases below the detection threshold of $0.02$~nm.
 \\ \indent
%**************************************************************************** 
Experiments performed in \cite{Zahra} suggest that the electron surface affinity of the microparticle material should not exceed that of the material of the QDs.
In this case, the surplus charge is concentrated in the QD layer and the local electric field is the subject for large fluctuations which result in measuarable Stark shifts.
To mitigate the unavoidable plasma-induced damage of the QDs, protective layers may be deposited above the layers of QDs.
In case of a thin ($\sim 1$~nm) protective layer made of a material with the negative electron surface affinity, the arrangement of the surplus electrons will closely correspond to the surface-surplus-charge assumption of the present model.
 \\ \indent
%**************************************************************************** 
Manufacturing technology of such microsensors as well as quantitative prediction of their lifetime in plasmas is out of the scope of the present work.
Those two issues as well as proof-of-principle experiments with the QD-based charge microsensors will be the subject of our further work.
 \\ \indent
%**************************************************************************** 
Development of QD-coated charge microsensors will allow to perform local measurements of their charges using visible-light optics. 
\label{Sec:Conc}
%**************************************************************************** 
\section{Acknowledgements}
Z. M. and J. B. appreciate the financial support from the Dutch Research Council (NWO), project number 15710.
The authors thank Prof. S. Chatterjee for helpful discussions and Dr. H.M. Thomas for careful reading of our manuscript.
 \\ \indent
%**************************************************************************** 

\end{document}